\documentstyle[12pt]{article}
\begin{document}
\begin{flushright}
DPKU9206 \\ BUTP-92/09  \\
April, 1992
\end{flushright}
\vspace{ .5cm}
\begin{center}
{\LARGE\bf  Neutrino Magnetic Moment Model and Solar Neutrino Problem}\\
\vspace{2 cm}
{\Large Yasuji Ono}\\
\vspace {0.5cm}
Department of Physics, Kanazawa University Kanazawa 920, Japan \\
\vspace {0.8cm}
{\Large and}\\
\vspace{ 0.8cm}
{\Large Daijiro Suematsu}\\
\vspace {0.5cm}
    Institute for Theoretical Physics, University of Bern,\\
     Sidlerstrasse 5, CH-3012 Bern,
Switzerland\footnote{Present address:
Department of Physics, Kanazawa University,
        Kanazawa 920, Japan.  }  \\
\end{center}
\vspace{2cm}
{\Large\bf Abstract}\\
We examine the time variation problem of solar neutrinos in the
spin-flavour precession mechanism taking into account the
$\nu_e$-$\nu_{\mu}$ mixing.
The models with the small and large mixing angle are studied.
It shows that the models which realize $m_{\nu_e\nu_e} \sim
m_{\nu_{\mu}\nu_{\mu}}
\ll m_{\nu_e\nu_{\mu}}$ and
$m_{\nu_e\nu_e} \sim m_{\nu_{\mu}\nu_{\mu}}
{^>_\sim} m_{\nu_e\nu_{\mu}}$ seem to have preferable time variation
features.
It is very interesting that the former type of models can give the large
magnetic moment to neutrinos and also
suppress the radiative mass correction naturally.

\newpage
{\it 1.Introduction}\\
The solar neutrino problem is one of the challenging problems in the recent
neutrino physics\cite{Pec}.
So called solar neutrino problem has two aspects.
One of them is the depletion of the solar neutrino flux in comparison with
that predicted from the
standard solar model (SSM)\cite{BU}.
This has been pointed out by the radio chemical $^{37}$Cl experiments of
Davis and his colaborators\cite{Dav}.
Recently it was confirmed by the water \v{C}erenkov detector in
Kamioka\cite{Hir}.
Another aspect is about the existence of the time variation of the solar
neutrino flux anti-correlated with the sunspot activity.
The $^{37}$Cl experiment shows this anti-correlation\cite{Dav}.
On the other hand the data of KamiokandeII experiment suggest that it is
 constant in time\cite{Hir}.
At a first sight there seems to be a discrepancy if we consider these results
seriously.

For the former problem there is a very simple solution called MSW\cite{Mik}
\cite{Kuo}, which uses
the resonance enhancement of the neutrino oscillation in the matter.
But this mechanism cannot explain any kind of time variation.
To explain the time variation problem, we need other mechanisms.
If neutrino has considerably large magnetic moment, neutrinos precess
into the sterile or other flavor neutrinos in the magnetic field in the sun
as suggested in some papers\cite{Cis}\cite{VVO}\cite{LM}\cite{Akh}.
This mechanism seems to be able to explain both aspects of the problem
except for the discrepancy about the time variation between two detectors.
However, it needs rather large extensions of the standard model
to make the neutrino magnetic moment large.
There are many trials toward the model construction\cite{MAG}\cite{BABUA}.
Simple minded extensions bring also the large mass correction and are ruled
out from the experimental neutrino mass bound.

In the previous work\cite{ONO} we showed the possibility that in the
spin-flavor precession mechanism there are some parameter regions which can
explain the difference in the time variation of solar neutrino between
two experiments.
In that explanation, essential points are that two experiments use the
different processes for the detection of neutrinos and also different energy
neutrinos have different precession features in the magnetic
field\cite{ONO}\cite{BABUB}.
The $^{37}{\rm Cl}$ detector is sensitive only to $\nu_e$ neutrinos
above 0.86MeV.
The average energy of $^8 {\rm B}$ neutrinos is $\sim$10MeV.
The total average energy of $^7 {\rm Be}$, $^{13} {\rm N}$,
$^{15} {\rm O}$ and pep neutrinos  is $\sim$1MeV.
Therefore all of them are included in this region.
On the other hand the threshold of KamiokandeII detector is 7.5MeV and
sensitive only to $^8{\rm B}$ neutrinos but it can react to $\bar\nu_e,
\nu_\mu , \bar\nu_\mu$ also.
These differences can produce the apparent discrepancy between two experimental
 results.

In this paper we extend this analysis into the case where
$\nu_e$-$\nu_{\mu}$ mixing exisits and study the relation between the
typical neutrino magnetic moment model and its time variation feature of
solar neutrino flux.
We suggest that if we take such a view point, we may have a
possibility to obtain the information
on the particle physics models which produce the large neutrino magnetic
moment on the basis of the solar neutrino experiments.\\

{\it 2.The mixing angle}\\
As is well-known, to give neutrinos the large magnetic moment of order
 $10^{-11}\mu_B$ which is an experimental upper bound,
we must extend the standard model somehow.
However, if we can make the neutrino magnetic moment
 large by a certain radiative correction, the same graph eliminated the
photon line which contributes to
 the magnetic moment brings in general the large mass correction to
neutrinos.
To forbid such a mass correction, it is usual to introduce a Lagrangian
symmetry ${\cal G}$ which may break down to its subgroup
${\cal H}$\cite{MAG}\cite{BABUA}.
In the most models ${\cal H}$ contains $L_e-L_{\mu}$ symmetry or its discrete
subgroup.
If ${\cal H}$ is exactly conserved, only off-diagonal mass appears and
there is no mass difference between
the neutrino flavors and then the solar neutrino problem cannot be
explained in the resonant spin-flavor precession\cite{LM}\cite{Akh}.
As the result we must break ${\cal H}$ softly to explain it.
In this situation taking account of the radiative mass correction, there
appears inevitably the extremely large $\nu_e$-$\nu_{\mu}$ mixing.

The mass matrices in the mass eigenstates $(\nu_1,\nu_2)$ and weak
eigenstates $(\nu_e,\nu_{\mu})$ are related as
\begin{equation}
  \left( \begin{array}{ll}
     m_1 & 0 \\  0 & m_2  \end{array} \right)
 = \left( \begin{array}{ll}
   \cos\theta & -\sin\theta \\   \sin\theta & \cos\theta
   \end{array} \right)
   \left( \begin{array}{ll}m_{\nu_e\nu_e} & m_{\nu_e\nu_{\mu}} \\
    m_{\nu_e\nu_{\mu}} &  m_{\nu_{\mu}\nu_{\mu}} \end{array} \right)
   \left( \begin{array}{ll}
   \cos\theta & \sin\theta  \\   -\sin\theta & \cos\theta  \end{array}
\right) .
\end{equation}
{}From this we get
\begin{equation}
 \Delta m^2 \equiv m_2^2-m_1^2
  =(m_{\nu_{\mu}\nu_{\mu}}+m_{\nu_e\nu_e})\sqrt{(m_{\nu_{\mu}\nu_{\mu}}
  -m_{\nu_e\nu_e})^2+4m_{\nu_e\nu_{\mu}}^2}, \end{equation}
and
\begin{equation}
 \tan 2\theta ={2m_{\nu_e\nu_{\mu}} \over m_{\nu_{\mu}\nu_{\mu}}-
m_{\nu_e\nu_e} }.  \end{equation}
In the model which has the restrictive symmetry, $m_{\nu_e\nu_e}$, $
m_{\nu_{\mu}\nu_{\mu}}$ and
 $m_{\nu_e\nu_{\mu}}$
can be considerably constrained.
As an example, we take the model proposed by Babu et al.\cite{BABUA}.
Their model has custodial $SU(2)$ as ${\cal H}$ which is softly broken.
Because of this symmetry, they get
$m_{\nu_{\mu}\nu_{\mu}} { ^>_\sim} m_{\nu_e\nu_e}
 \sim10^{-7}$eV and
 $m_{\nu_e\nu_{\mu}} \sim 1$eV.
This realizes $\Delta m^2  {^<_\sim}  O\bigl (10^{-7})$ eV$^2$ which is
needed to explain the solar neutrino deficit due to the resonant spin-flavor
precession.
But the above value gives $\tan 2\theta \sim 10^{7}$.
This feature seems to be general in the models which have the softly broken
 $L_e-L_{\mu}$ symmetry or its discrete subgroup as the subgroup of $
{\cal H}$.
It is a non-trivial problem what time variation behavior the models with
non-zero $\sin 2\theta$, in particular, $\sin 2\theta \cong 1$ show even
if $\Delta m^2  {^<_\sim}  O\bigl (10^{-7})$ eV$^2$.
If we relate this to the above mentioned time variation problem and analyze
 them, we may find some informations about the features of the various
models of large transition magnetic moment.
This is the subject in the following arguments.\\

{\it 3. Time evolution of the solar neutrino}\\
We consider the solar neutrino problem in the framework of the spin-flavor
precession mechanism with non-zero $\nu_e$-$\nu_{\mu}$ mixing angle
$\theta$.
As discussed in the previous part, we should note that this is the
general situation in the model with large transition magnetic moment.
Here we restrict our study to two flavors case.

The time evolution equation of the neutrinos in the magnetic field $B$ is
 expressed in the weak interaction basis as
\begin{equation}
 i{d \over dt}\left(  \begin{array}{l}
   \nu_e \\ \nu_\mu \\ \bar\nu_e \\  \bar\nu_{\mu} \end{array} \right)
  =\left( \begin{array}{llll} a_e &
   {\Delta m^2 \over 4E}\sin 2\theta & 0 &B\mu \\
  {\Delta m^2 \over 4E}\sin 2\theta &
  {\Delta m^2 \over 2E}\cos 2\theta +a_{\mu} & -B\mu & 0 \\
     0 & -B\mu &  -a_e &
    {\Delta m^2 \over 4E}\sin 2\theta \\
     B\mu & 0 & {\Delta m^2 \over 4E}\sin 2\theta &
     {\Delta m^2 \over 2E}\cos 2\theta -a_{\mu}
   \end{array} \right)
     \left( \begin{array}{l} \nu_e \\ \nu_\mu \\ \bar\nu_e \\
                                           \bar\nu_{\mu} \end{array}
     \right).  \end{equation}
$a_e$ and $a_{\mu}$ are the coherent scattering effect of neutrinos
with matter and expressed as
\begin{equation}  a_e=\sqrt 2 G_F(n_e-{1 \over 2}n_n), \qquad
  a_{\mu}={1 \over \sqrt 2}G_F(-n_n).  \end{equation}
$G_F$ is Fermi coupling constant.
$n_e$ and $n_n$ are the electron and neutron number densities in the sun,
respectively.\footnote{In the following
study we use $n_n\simeq {1 \over 6}n_e$ in the sun.}
$n_e$ is known to be approximately represented \cite{BU} as
\begin{equation}
n_e=2.4\times 10^{26}\exp (-r/0.09R_{\odot}) \quad cm^{-3}. \end{equation}
Since we consider the neutrino relativistic, we take $r=t$.
In our discussion
we assume that $B$ is constant throughout the sun and $B\mu \sim 10^{-16+y}$eV.
This corresponds approximately to $\mu \sim 10^{-11}\mu_B$ and $B \sim
10^{4+y}$G.
$y$ represents the change of the magnetic field and we take $0<y<3$.

We are interesting in the time variation feature of the solar neutrino flux
, particularly, and then for a while we
focus our attention on $(\nu_e, \bar\nu_{\mu})$ sector of eq.(4),
\begin{equation}
 i{d \over dt}\left(  \begin{array}{l}
   \nu_e \\   \bar\nu_{\mu} \end{array} \right)
  =\left( \begin{array}{ll} a_e &B\mu \\
         B\mu &    {\Delta m^2 \over 2E}\cos 2\theta -a_{\mu}
   \end{array} \right)
     \left( \begin{array}{l} \nu_e \\   \bar\nu_{\mu} \end{array}
     \right).  \end{equation}
There are two possibilities for the large $\nu_e \rightarrow
 \bar\nu_{\mu}$ transition.
One is the resonant spin-flavor precession in the sun and the
other case has no resonance in the sun but has the large precession because of
$ B\mu >{\Delta m^2 \over 2E}\cos 2\theta -a_{\mu}-a_e $.

At first we consider the former possibility.
As suggested in refs.\cite{LM}\cite{Akh}, this precession $\nu_e \rightarrow
 \bar\nu_{\mu}$
has the resonance structure.
The resonance occurs at
${\Delta m^2 \over 2E}\cos 2{\theta}=a_e +a_{\mu}.$
If we require this resonance point to exist in the convection zone
$(0.7R_\odot \leq
r \leq R_\odot )$, ${\Delta m^2 \over 2E}\cos 2{\theta}$ should satisfy
\begin{equation}
4.0 \times 10^{-16}{\rm eV} \leq {\Delta m^2 \over 2E}\cos 2{\theta}
\leq 1.1 \times 10^{-14}{\rm eV}.
\end{equation}
This condition is necessary to be satisfied to explain the sufficient
time variation of solar neutrinos anti-correlated with sunspot nembers.
To see the qualitative feature of $\nu_e \rightarrow
 \bar\nu_{\mu}$ transition
 we solve eq.(7) around the resonance point following Landau and
Zener\cite{Lan}.
The transition amplitude is expressed as
\begin{equation} P_x =\exp [-2\pi\gamma (1-\tan^2\beta_{N=0} )] \end{equation}
where
\begin{equation} \gamma ={ (B\mu )^2 \over
             {\Delta m^2 \over 2E}\cos 2\theta \vert {1 \over N}
                         {dN \over dt} \vert_{t_r} },
\qquad N=a_e+a_{\mu}.
\end{equation}
 $\beta$ is the angle which is used to diagonalize the matrix
in the right-hand side of eq.(7) and
\begin{equation} \tan 2\beta = {2B\mu \over
                  {\Delta m^2 \over 2E}\cos 2\theta - N }.
 \end{equation}
Here we introduced the correction $1-\tan^2\beta_{N=0}$ which is due to the
specific electron density function we used here\cite{Kuo}.
After substitution of eqs.(10) and (11) into eq.(9), we get
\begin{equation}
P_x = \exp [-\pi { (2B\mu )^2
\over ({\Delta m^2 \over 2E}\cos 2\theta +\sqrt{({\Delta m^2 \over 2E}\cos
2\theta )^2 +( 2B\mu )^2 })
\vert {1 \over N}{dN \over dt}\vert_{t_r}} ].  \end{equation}
For the electron number density (6) we have
$\vert {1 \over N}{dN \over dt}\vert_{t_r} \sim 3.2 \times 10^{-15}$eV.
Following the procedure of Parke\cite{Par}, we can express the probability
 $P_s$ such that
$\nu_{e}$ remains $\nu_{e}$ on the surface of the sun as follows
\begin{equation}
 P_s = {1 \over 2} +({1 \over 2}-P_x)\cos 2\beta_1\cos 2\beta_2.
                                 \end{equation}
$\beta_1$ and $\beta_2$ are the values of $\beta$ at the core and the surface
of the sun.
The value of $\beta $ can generally change from $\beta_1 \sim {\pi \over 2}$
to $
\beta_2(<{\pi \over 4})$ when the neutrino travels from the core to the
surface of the sun through the resonance point.
Here it should be noted that $P_s$ has the neutrino energy dependence as shown
in eq.(12).

Next we consider the case in which no resonance is in the sun.
To see the qualitative feature we consider that $a_e$ and $a_\mu$ are
constants for a while.
If we put $ A={\Delta m^2 \over 2E}\cos 2\theta -N $ and solve
eq.(7), we get
\begin{equation}
 P_s=1 - {(2B\mu )^2 \over A^2 +(2B\mu )^2}
\sin^2 ({ (A^2 +(2B\mu )^2)^{1 \over 2} \over 2}R).
\end{equation}
For an effective $\nu_e \rightarrow \nu_\mu $ transition $B\mu >\vert A\vert $
should
be satisfied and then the neutrino energy dependence in $P_s$ becomes very
small.
Here $R$ corresponds to the distance where such a condition is satisfied in the
 sun.

In both cases we need to take account of the vacuum oscillation on the way from
the sun to the
earth due to the non-zero mixing angle $\theta$.
 The $\nu_e$ preserving
probability on the earth is expressed as
\begin{equation}
 P_e = \bar P_s +{1 \over 2}[(1-2\bar P_s)\sin^22\theta -
            \sqrt{\bar P_s(1-\bar P_s)}\sin 4\theta ],  \end{equation}
if the averaging procedure of oscillation parts is justified
 (${\Delta m^2 \over 4E}D>\pi$).
$D$ is the distance from the surface of the sun to the earth
$(\simeq 1.5 \times 10^8 km)$.
$\bar P_s$ stands for the value in which the MSW transition effect is
also considered.\\

{\it 4. Time variation features}\\
We proceed to the study of time variation features in the various models.
Before it, it is useful to note that there exists the interesting situation
where
there is no time variation of $P_s$ associated with the change of magnetic
field even if the neutrino has the large magnetic moment.
In eq.(13) this is related to the adiabaticity$(P_x=0)$ of the spin-flavor
precession\cite{ONO}\cite{BABUB}.
If $\cos 2\beta_2 \sim 0$ is realized in this adiabatic situation, eq.(13)
shows
$P_s \cong {1 \over 2}$.
On the other hand in eq.(14) it occurs as the averaging effect when $B\mu R
> \pi(B\mu > \vert A\vert )$ and $P_s \cong {1 \over 2}$.
These mean that the neutrino large magnetic moment does not necessarily
induce the time variation of the solar neutrino flux according to the change of
the magnetic field of the sun.
Thus the determination of the parameter range where the time variation becomes
large is the crucial point to explain the solar neutrino problem in the
spin-flavor precession framework.

In order to study the relation between the neutrino magnetic moment models
and the solar neutrino time variation, it is very instructive to consider
the two extreme cases, that is, $\sin 2\theta \sim 0$ and $\cong 1$.\\
\indent (i) $\sin 2\theta \sim 0$ case.\\
In this case the main transition of $\nu_e$ occurs almost through
$\nu_e \rightarrow \bar\nu_\mu$.
Equations(11) and (12) become
\begin{equation} \tan 2\beta \sim {2B\mu \over
                  {\Delta m^2 \over 2E} - N },
 \end{equation}
\begin{equation} P_x \sim \exp [-\pi {(2B\mu )^2 \over ({\Delta m^2 \over 2E} +
   \sqrt{({\Delta m^2 \over 2E})^2 + (2B\mu )^2}) \vert {1 \over N}{dN \over
dt}
\vert_{t_r} }].  \end{equation}
Because of $\sin 2\theta \sim 0$, there is no vacuum oscillation and $P_e$
in eq.(15) is equivalent to the value on the surface of the sun
$ P_e \cong P_s$.
The resonance condition (8) is now
$4.0 \times 10^{-16}{\rm eV} {^<_\sim} {\Delta m^2 \over 2E}
{^<_\sim} 1.1 \times 10^{-14}{\rm eV}$.
As a typical $\Delta m^2$ value, let's take $\Delta m^2=2 \times 10^{-
8}$eV.
Then ${\Delta m^2 \over 2E}= 10^{-14}{\rm eV}\ (E=1{\rm MeV})$ and
 ${\Delta m^2 \over 2E}= 10^{-15}{\rm eV}\ (E=10{\rm MeV})$.
As easily checked by eqs.(16) and (17), these values
show that for 1MeV neutrinos $P_x$ is nonadiabatic and can
have rather large $B\mu$ dependence in a certain $B\mu$ range but for 10MeV
neutrinos $P_x$ is nearly
 adiabatic and depends on $B\mu$ weakly in comparison with 1MeV neutrinos
(Fig.1).
Moreover the very small $\cos 2\beta_2 $ is realized for 10MeV neutrinos but it
is very large for 1MeV neutrinos (Fig.2).
As a result, in this case we can find parameter regions where the low energy
neutrinos generally show rather
larger time variation than the high energy neutrinos (Fig.3).
Numerical studies have been done in $\sin 2\theta =0$ case and very
interesting results have been obtained for the time variation
problem\cite{ONO}\cite{BABUB} as predicted from above qualitative
arguments.
In ref.\cite{BABUB}  $\sin 2\theta =0.1$ case has also been studied.
There it is shown that there is no essential difference from $\sin 2\theta
=0$.
This case corresponds to the model with $m_{\nu_e\nu_e},
m_{\nu_{\mu}\nu_{\mu} } \gg m_{\nu_e\nu_{\mu} }$
$(\Delta m^2 \sim m_{\nu_e\nu_e}^2-m_{\nu_{\mu}\nu_{\mu}}^2)$.
However, within our knowledge this type of model with both of the large
transition magnetic moment and the natural
 suppression of the radiative mass correction
has not been proposed by now.\\
\indent (ii) $\sin 2\theta \cong 1$ case.\\
In this case, from eq.(4)
 the neutrino flux can be expected to be sufficiently reduced
$(\bar P_s < {1 \over 2})$
due to both the large mixing and the large magnetic moment effect
as in pseudo Dirac case\cite{KOBA}.
And also it seems
to be possible to exhibit the time variation unless $B\mu$ becomes too
large and the adiabaticity or the averaging effect appears.
Moreover, in the almost maximal mixing case there is an interesting feature.
Kamiokande detector can interact with all of $\nu_e, \nu_\mu , \bar\nu_e,
\bar\nu_\mu$ in contrast with $^{37}$Cl detector which reacts only to
$\nu_e$.
These cross sections for 10MeV neutrinos are\cite{MORI}
$\sigma (\bar\nu_e e)=0.42\sigma (\nu_e e),
 \sigma (\nu_\mu e)=0.17\sigma (\nu_e e),
\sigma (\bar\nu_\mu e)=0.14\sigma (\nu_e e).$
As pointed out in ref.\cite{KOBA}, in the case of almost maximal mixing,
independently of the initial values on the surface of the sun,\footnote{For
instance this can be easily seen also from eq.(15).}
the vacuum
oscillation from the sun to the earth causes the probabilities $a, a,
{1-2a \over 2}, {1-2a \over 2}$ for $\nu_e, \nu_\mu , \bar\nu_e,
\bar\nu_\mu$, respectively.\footnote{As pointed out in
refs.\cite{MORI}\cite{BARB}, the fraction of $\bar \nu_e$ may be constrained
from the data on the background of the KamiokandeII experiment.
Although it may be crucial factor to select the neutrino large magnetic moment
model, in our following analysis we do not take account of this constraint.}
Thus the detection ratio of two detectors are
$$P_{Cl}=a, \qquad P_{K-II}=0.61a +0.28.$$
This means that $0.2<P_{Cl}<0.56$ when $0.4<P_{KII}<0.6$.
We can expect this amount of depletion and variation through the cooperation
of  the almost maximal mixing and $B\mu$ effect as the psuedo Dirac case in
ref.\cite{KOBA}.

Now we examine the relation between the models with $\sin 2\theta \cong 1$ and
time variation features
in more detail.
At first we consider the case in which the resonance exists in the sun.
In this case the time variation can be expected through the resonance effect.
But the neutrino energy dependence of $P_s$ in eq.(12) disappears in contrast
with the (i) case.
 The difference of the time variation between two experiment must be explained
only by the difference of the processes used in the detection.
We assume $\cos 2\theta \sim 10^{-\alpha}$ $ (\alpha >1)$.
{}From the resonance condition (8),
$8.0 \times 10^{-9+\alpha}{\rm eV}^2 {^<_\sim} \Delta m^2
{^<_\sim} 2.2 \times 10^{-7+\alpha }{\rm eV}^2$ for 10MeV neutrinos.
Following eqs.(2) and (3), we get ${2m_{\nu_e\nu_\mu} \over m_{\nu_\mu\nu_\mu}-
m_{\nu_e\nu_e}} \sim 10^\alpha $ and $m_{\nu_\mu\nu_\mu}^2-
m_{\nu_e\nu_e}^2 \sim 10^{-8}$ in the case of $\Delta m^2 \sim
10^{-8+\alpha}$.
We must impose electron neutrino mass bound on $m_{\nu_e\nu_\mu} {^<_\sim}
10$eV.
This results in $m_{\nu_\mu\nu_\mu}-m_{\nu_e\nu_e} {^<_\sim} 10^{1-\alpha}$ and
$m_{\nu_\mu\nu_\mu}+m_{\nu_e\nu_e} {^>_\sim} 10^{-7+\alpha}$.
To satisfy these simultaneously, $\alpha {^>_\sim} 4$ is required.
If $\alpha$ become large, the extreme fine tuning between
$m_{\nu_\mu\nu_\mu}$ and $m_{\nu_e\nu_e}$ is necessary.
{}From this fact, $m_{\nu_\mu\nu_\mu} \sim m_{\nu_e\nu_e}\sim 10^{-4}$eV and
$m_{\nu_e\nu_\mu}\sim 1$eV is considered as a natural possibility.
However, as predicted from the pseudo Dirac case\cite{KOBA}, when $\Delta
m^2 {^>_\sim} 10^{-7}$eV the transition $\nu_e \rightarrow \nu_\mu$ due to
the maximal mixing effect becomes adiabatic and as a result we will not get
the sufficient suppression of the flux.
This case seems not to be so attractive.

Next we consider the case in which no resonance exists in the sun.
{}From eq.(8) this is realized when ${\Delta m^2 \over 2E}\cos 2\theta
{^<_\sim} 10^{-16}$.
No resonance is in the convection zone in the sun but  $B\mu > N $ is satisfied
in the convection zone.
If $B\mu R < \pi$ is satisfied, the change of magnetic field will be able to
induce the sufficient $\nu_e$ flux change.
In fact as we found numerically in ref.\cite{ONO}, there is almost no
dependence on ${\Delta m^2 \over 2E}\cos 2\theta$ in the
form of $P_s$ derived from eq.(7) when ${\Delta m^2 \over 2E}\cos 2\theta
< 10^{-16}$eV.
Thus we can apply that result in this case because of $\vert A \vert \sim N$.
Following it, the sufficient time variation (from 1 to 0.3) of $P_s$ can be
expected associated with the order two change of the magnetic field $(10^{-
16}{\rm eV} {^<_\sim} B\mu {^<_\sim} 10^{-14}{\rm eV})$.
This seems to be very preferable feature.
We can use this change of
magnetic field fully to explain the time variation of solar neutrino because
the flux depletion is the combined effect of
the mixing and $B\mu$ in this case.
As an example, in Fig.4 we give the numerical results in $\sin 2\theta =1$
case.We can easily see that our above predictions are realized in it.
In the resonant solution case where $\sin 2\theta =0$, the necessary change of
the magnetic field is only factor 2 or 3 and much smaller than the expected
value.\cite{ONO}
This is because $\nu_e \rightarrow \bar\nu_\mu$ transition is the only
origin of the
depletion of the neutrino flux.
The neutrino energy independence in  $P_s$ does not seem to be matter for the
explanation of the difference between two experiments because of eq.(18).
We should note that the model of ref.\cite{BABUA} and other models with
similar kind of symmetries correspond to this interesting category of models.
 This model is realized naturally by
$m_{\nu_e\nu_\mu} \gg m_{\nu_e\nu_e} \sim m_{\nu_\mu\nu_\mu}$ without fine
tunings.

Here we comment a bit on the maximal mixing case ($\sin 2\theta =1$).
It can be realized in two ways, that is, $m_{\nu_\mu\nu_\mu}= m_{\nu_e\nu_e}
\not= 0 (\Delta m^2 \not=0)$ and  $m_{\nu_\mu\nu_\mu}= m_{\nu_e\nu_e}=0
(\Delta m^2 =0)$.
The former case corresponds to the softly broken ${\cal H}$ type model and
has the similar time variation property as $\sin 2\theta \cong 1$.
On the other hand the latter case which corresponds to the exact ${\cal H}$
symmetry has no $\nu_e \rightarrow \nu_\mu$ transition.
It is similar to $\sin 2\theta \sim 0$ case except for having no
resonance.\footnote{It should be noted that there might be a possibility to
explain the solar neutrino problem even if $\Delta m^2 =0$. }
No resonance means that $P_s$ has no neutrino energy dependence.
Thus the ratio of the time variation between $^{37}$Cl and KamiokandeII
experiments will
become smaller than that of $\sin 2\theta \sim 0$ case.
The softly broken $\cal H$ is favorable for the explanation of the time
variation problem also in the maximal mixing case.\\

{\it 5. Conclusions}\\
{}From the study of extreme cases we found that the different $\theta$ models
show the considerably different time variation of the solar neutrino
flux due to the different reasons.
And also our study suggests that only models with rather restricted $\sin
2\theta$ value
are natural as the large neutrino magnetic moment models and also can
produce the sufficient time variation of solar neutrino flux for a certain
parameter range.

The tipical models which give the large transition magnetic moment have
the symmetry ${\cal H}$ which contains $L_e-L_{\mu}$ symmetry or its discrete
subgroup.
 These symmetries are essential in these models to suppress the large mass
correction and the rare process like $\mu \rightarrow e\gamma,\  \mu
\rightarrow ee\bar e$ and so on.
On the other hand these symmetries lead to $\sin 2\theta \cong 1$ in
general.
 Fotunately, from the above analysis we conclude that the softly broken
${\cal H}$ type model can induce the certain flux depletion
and also can cause the different time
variation results between two experiments associated with the change of
magnetic field.
The models with $\sin 2\theta \sim 0$ have the same property but in such models
the natural supperssion of the radiative mass correction is not known.

In summary, we investigated the time variation problem of the solar neutrino
 flux in the framework of the spin-flavor precession in the
magnetic field.
In the models which give neutrinos the large transition magnetic moment
there exists the non-zero $\nu_e$-$\nu_{\mu}$ mixing angle $\theta$
generally.
Taking account of this mixing we analyzed the solar neutrino
experimental results of $^{37}$Cl and KamiokandeII.
As a result we found that models with $\sin 2\theta \cong 1$ for the large
 neutrino transition magnetic moment proposed by now seem to have a very
interesting feature to
 induce the time variation of solar neutrino flux.
This suggests that
the models which realize $m_{\nu_e\nu_e}
 \sim m_{\nu_{\mu}\nu_{\mu}} \ll   m_{\nu_e\nu_{\mu}}$
 $\ (\sin 2\theta \cong 1)$
are promising to explain the time variation
of the solar neutrino flux.
The $\sin2 \theta \sim 0$ model which is realized by
$m_{\nu_e\nu_e}
 \sim m_{\nu_{\mu}\nu_{\mu}}  {^>_\sim}   m_{\nu_e\nu_{\mu}}$
 $\ (\sin 2\theta \sim 0)$ is also promising.
However, it is very challenging problem how to construct a small
$\sin 2\theta$ model which realize simultaneously the large transition magnetic
moment and the natural suppression of the large mass correction.
More detailed numerical investigation will be needed for more quantitative
arguments.\\

We  would like to thank the members of particle physics group of Kanazawa
University for valuable discussions.
One of the author(D.S.) is grateful for the hospitality of ITP of
University of Bern.
D.S. was partially supported by Swiss National Science Foundation and Japan
Society for the Promotion of Science.
\newpage

\newpage


\begin{thebibliography}{99}
\bibitem{Pec}For solar neutrino problem, see for example,\\
  R.~D.~Peccei, DESY 89-043, UCLA/89/TEP/12.\\
   J.~N.~Bahcall, "Neutrino Astrophysics"
 Cambridge University Press 1989.
 %
\bibitem{BU}J.~N.~Bahcall and R.~K.~Ulrich, Rev. Mod. Phys. {\bf 60} (1988)
297.
%
\bibitem{Dav}R.~Davis,~Jr., in Proc. of
7th Workshop on Grand Unification, Toyama, Japan, ed. by J.~Afafune (World
Scientific, Singapore, 1987) p237;
 in Neutrino '88 Proc. of 13th Int. Conf. on Neutrino Physics,
Boston, 1988, ed. by J.~Schneps {\it et al.} (World Scientific, Singapore,
1988) p502.
%
\bibitem{Hir}K.~S.~Hirata {\it et al.}, Phys. Rev. Lett.{\bf 61} (1988) 2653;
  Phys.
Rev. Lett. {\bf 65} (1990) 1297;
 Phys. Rev. Lett. {\bf 65} (1990) 1301.
%
\bibitem{Mik}S.~P.~Mikheyev and A.~Yu~Smirnov, Sov. J. Nucl. Phys. {\bf 42}
 (1985) 913;\\
    L.~Wolfenstein, Phys. Rev. {\bf D17} (1978) 2369;
     Phys. Rev. {\bf D20} (1979) 2634.
%
\bibitem{Kuo}I.~K.~Kuo and J.~Pantaleone, Rev. Mod. Phys, {\bf 61} (1989) 937.
%
\bibitem{Cis}A.~Cisneros, Astrophys. Space Sci. {\bf 10} (1981) 87.
%
\bibitem{VVO}M.~B.~Voloshin and M.~I.~Vysotsky, ITEP Report No.1,
1986;\\
 L.~B.~Okun, Yad. Fiz. {\bf 44} (1986) 847
 [Sov. J. Nucl. Phys. {\bf 44} (1986) 546];\\
 M.~B.~Voloshin, M.~I.~Vysotsky and L.~B.~Okun,
 Zh. Eksp. Teor. Fiz. {\bf 91} (1986) 754
 [Sov. Phys. JETP {\bf 64} (1986) 446].
%
 \bibitem{LM}C.~S.~Lim and W.~J.~Marciano, Phys. Rev.
{\bf D37} (1988) 1368.
%
\bibitem{Akh}E.~Kh.~Akhmedov, Phys. Lett. {\bf B213} (1988) 64.
%
\bibitem{MAG}M.~Fukugita and T.~Yanagida, Phys. Rev. Lett. {\bf 58} (1987)
1807,\\
M.~B.~Voloshin, Yad Fiz. {\bf 48} (1988) 804
[Sov. J. Nucl. Phys. {\bf 48} (1988) 512],\\
R.~Barbieri and R.~N.~Mohapatra, Phys. Lett. {\bf B218} (1989) 225,\\
K.~S.~Babu and R.~N.~Mohapatra, Phys. Rev. Lett. {\bf 63} (1989) 228;
 {\bf 64} (1989) 1705,\\
M.~Leurer and N.~Marcus, Phys. Lett. {\bf B237} (1990) 81,\\
S.~Barr, E.~M.~Freire and A.~Zee, Phys. Rev. Lett. {\bf 65} (1990) 2626,\\
D.~Chang, W.~-Y.~Keung and G.~Senjanovic, Phys. Rev. {\bf D42} (1990)
1599,\\
H.~Georgi and L.~Randall, Phys. Lett. {\bf B244} (1990) 196.
%
\bibitem{BABUA}K.~S.~Babu and R.~N.~Mohapatra, Phys. Rev. {\bf D43}
(1991) 2278.
%
\bibitem{ONO}Y.~Ono and D.~Suematsu, Phys. Lett. {\bf B271} (1991) 165.
%
\bibitem{BABUB}K.~S.~Babu, R.~N.~Mohapatra and I.~Z.~Rothstein, Phys. Rev.
{\bf D44} (1991) 2265.
%
\bibitem{Lan}L.~Landau, Phys. Z. USSR {\bf 2} (1932) 46; \\
C.~Zener, Proc. Roy. Soc. {\bf A137} (1932) 696.
%
\bibitem{Par}S.~J.~Parke, Phys. Rev. Lett. {\bf 57} (1986) 1275.
%
\bibitem{KOBA}M.~Kobayashi, C.~S.~Lim and M.~M.~Nojiri, Phys. Rev. Lett.
{\bf 67}
 (1991) 1685.
%
\bibitem{MORI}C.~S.~Lim, M.~Mori, Y.~Oyama and A.~Suzuki, Phys. Lett.
{\bf B243 } (1990) 389.
%
\bibitem{BARB}R.~Barbieri, G.~Fiorentini, G.~Mezzorani and M.~Moretti, Phys.
Lett. {\bf B259} (1991) 119.
\bibitem{BARG}V.~Barger, N.~Deshpande, P.~B.~Pal, R.~J.~N.~Phillips and
K.~Whisnant, Phys. Rev. {\bf D43} (1991) R1759.
\end{thebibliography}
\end{document}